\documentstyle{article}
   \newcommand\?[1]{} % comments
\def\be{\begin{equation}}   \def\ee{\end{equation}}   \def\CR{$$ $$}
\def\IP{{\bf P}}    %{\hbox{\rm I\kern -1.6pt{\rm P}}}
\def\IL{{\bf L}}    %{\hbox{\rm I\kern -1.6pt{\rm L}}}
\def\IC{{\hbox{\rm C\kern-.58em{\raise.53ex\hbox{$\scriptscriptstyle|$}}
    \kern-.55em{\raise.53ex\hbox{$\scriptscriptstyle|$}} }}}
\def\IN{\hbox{I\kern-.2em\hbox{N}}}
\def\IR{\hbox{\rm I\kern-.2em\hbox{\rm R}}}
\def\IZ{\hbox{{\rm Z}\kern-.3em{\rm Z}}}
\def\IT{\hbox{\rm T\kern-.38em{\raise.415ex\hbox{$\scriptstyle|$}} }}
\def\notsub{\hbox{$\subset$\kern-.55em\hbox{/}}}

\def\n{\noindent}                  \def\txt#1{{\rm #1}}
\def\var{\, {\rm var}}   \def\Osc{\, {\rm Osc}}  
                    
                         \def\mod1{\,({\rm mod\ } 1)\,}
\def\ep{\varepsilon}     \def\phi{\varphi}       \def\la{\lambda}
           \def\IQ{{\bf Q}}
\def\1#1{\hbox{\large \bf 1}_{#1}}     % unit function on #1
\def\I|#1{{\big|}_{#1}}                % restriction to #1

\def\epd#1{\if #11 \ep^{-d} \else \ep^{-#1d} \fi}  % \epsilon^{-#1d}
\def\proof{\smallskip \noindent {\bf Proof. \ }}
\newcommand\filledsquare{\ \vrule width 1.5ex height 1.2ex}  %filled square
\def\qed{\hfill\filledsquare\linebreak\smallskip\par}

\newcommand\Lbrace{\left\lbrace}        \newcommand\Rbrace{\right\rbrace}
\newcommand\Lpar{\left(}                \newcommand\Rpar{\right)}
\newcommand\farrow[1]{\buildrel #1 \over \longrightarrow}

\newcommand\eqrefB[1]{(\ref{#1})} % to refer to equations

\newtheorem{theorem}{Theorem}[section]   %Numbering: Theorem--Other section
\newtheorem{lemma}{Lemma}[section]       %{lemma}[theorem]{Lemma}   section
\newtheorem{corollary}[lemma]{Corollary}      %[theorem]
\newtheorem{remark}[lemma]{Remark}            %[theorem]
\newtheorem{definition}[lemma]{Definition}    %[theorem]
             \catcode`@=11 \@addtoreset{equation}{section} \catcode`@=12

\newcommand\mlbscale{1pt} %to change the scale: \renewcommand\mlbscale{1.3pt}

\def\bfig(#1,#2)#3#4{\begin{figure} \begin{center}
    \framebox{\setlength{\unitlength}{\mlbscale} \begin{picture}(#1,#2) #3
    \end{picture} } \end{center} \caption{#4} \end{figure}}
\def\Bfig(#1,#2)#3#4{\begin{figure} \begin{center}
    \setlength{\unitlength}{\mlbscale} \begin{picture}(#1,#2) #3
    \end{picture} \end{center} \caption{#4} \end{figure}}
\def\bpic(#1,#2)#3{\setlength{\unitlength}{\mlbscale}
    \begin{picture}(#1,#2) #3 \end{picture}}
\def\picsquare{\hskip -0.5\@wholewidth%
    \vrule height \@halfwidth depth \@halfwidth width \@wholewidth}
\makeatletter \newcount\@x@diff \newcount\@y@diff \newdimen\x@diff
\newdimen\y@diff \newcount\num@segments \newcount\num@segmentsi
\newif\if@flippedargs
\def\lineslope(#1,#2){
 \ifdim #1 <0pt \@xdim= -#1 \else\@xdim=#1\fi
 \ifdim #2 <0pt \@ydim= -#2 \else\@ydim=#2\fi
 \ifdim\@xdim >\@ydim \@tempdima=\@xdim \@xdim=\@ydim \@ydim=\@tempdima
 \@flippedargstrue\else\@flippedargsfalse\fi% x < y
 \ifdim\@ydim >1pt \@tempcnta=\@ydim
             \divide\@tempcnta by 65536% now \@tempcnta=integral part of #1.
             \divide\@xdim \@tempcnta\fi
 \ifdim\@xdim <.083333pt \@xarg=1 \@yarg=0
  \else\ifdim\@xdim <.183333pt    \@xarg=6 \@yarg=1
  \else\ifdim\@xdim <.225pt       \@xarg=5 \@yarg=1
  \else\ifdim\@xdim <.291666pt    \@xarg=4 \@yarg=1
  \else\ifdim\@xdim <.366666pt    \@xarg=3 \@yarg=1
  \else\ifdim\@xdim <.45pt        \@xarg=5 \@yarg=2
  \else\ifdim\@xdim <.55pt        \@xarg=2 \@yarg=1
  \else\ifdim\@xdim <.633333pt    \@xarg=5 \@yarg=3
  \else\ifdim\@xdim <.708333pt    \@xarg=3 \@yarg=2
  \else\ifdim\@xdim <.775pt       \@xarg=4 \@yarg=3
  \else\ifdim\@xdim <.816666pt    \@xarg=5 \@yarg=4
  \else\ifdim\@xdim <.916666pt    \@xarg=6 \@yarg=5
       \else                      \@xarg=1 \@yarg=1%
 \fi\fi\fi\fi\fi\fi\fi\fi\fi\fi\fi\fi
 \if@flippedargs\relax\else\@tempcnta=\@xarg \@xarg=\@yarg
                           \@yarg=\@tempcnta\fi
 \ifdim #1 <0pt \@xarg= -\@xarg\fi  \ifdim #2 <0pt \@yarg= -\@yarg\fi
}

\newif\if@toosmall \newif\if@drawit \newif\if@horvline \def\drawlinestretch{0}
\def\drawline{\@ifnextchar [{\@idrawline}{\@idrawline[\drawlinestretch]}}
\def\@idrawline[#1](#2,#3){\@ifnextchar ({\@iidrawline[#1](#2,#3)}{\relax}}
\def\@iidrawline[#1](#2,#3)(#4,#5){\@drawline[#1](#2,#3)(#4,#5)
\@idrawline[#1](#4,#5)}

\def\@drawline[#1](#2,#3)(#4,#5){{%
\x@diff=#4\unitlength \advance\x@diff by -#2\unitlength
\y@diff=#5\unitlength \advance\y@diff by -#3\unitlength

\ifx\@linefnt\tenln \linethickness{0.5pt} \else \linethickness{0.9pt}\fi
\lineslope(\x@diff,\y@diff)% returns the two integers in \@xarg & \@yarg.

\@toosmalltrue
{\ifdim\x@diff <\z@ \x@diff=-\x@diff\fi
 \ifdim\y@diff <\z@ \y@diff=-\y@diff\fi
 \ifdim\x@diff >10pt \global\@toosmallfalse\fi
 \ifdim\y@diff >10pt \global\@toosmallfalse\fi}

\@drawitfalse\@horvlinefalse \ifnum#1 <0 \relax\else\@horvlinetrue\fi
\if@toosmall\@horvlinetrue\fi
\if@horvline
 \ifdim\x@diff =0pt \put(#2,#3){\ifdim\y@diff >0pt \@linelen=\y@diff \@upline
                                 \else\@linelen=-\y@diff \@downline\fi}%
 \else\ifdim\y@diff =0pt
          \ifdim\x@diff >0pt \put(#2,#3){\vrule \@height \@halfwidth \@depth
                                \@halfwidth \@width \x@diff}
                \else \put(#4,#5){\vrule \@height \@halfwidth \@depth
                                \@halfwidth \@width -\x@diff}\fi
       \else\@drawittrue\fi\fi % construct the line explicitly
\else\@drawittrue\fi

\if@drawit
\ifnum\@xarg< 0 \@negargtrue\else\@negargfalse\fi
\ifnum\@xarg =0 \setbox\@linechar%
\hbox{\hskip -\@halfwidth \vrule \@width \@wholewidth \@height 10.2pt
 \@depth \z@}
\else \ifnum\@yarg =0 \setbox\@linechar%
\hbox{\vrule \@height \@halfwidth \@depth \@halfwidth \@width 10.2pt}
\else \if@negarg \@xarg -\@xarg \@yyarg -\@yarg
      \else \@yyarg \@yarg\fi
\ifnum\@yyarg >0 \@tempcnta\@yyarg \else \@tempcnta -\@yyarg\fi
\setbox\@linechar\hbox{\@linefnt\@getlinechar(\@xarg,\@yyarg)}%
\fi\fi

\if@toosmall % => it isn't a horiz or vert line and is toosmall.
  \@dottedline[\picsquare]{.98\@wholewidth}%
(#2\unitlength,#3\unitlength)(#4\unitlength,#5\unitlength)%
\else

\ifnum\@xarg=0\relax\else\ifdim\x@diff >\z@ \advance\x@diff -\wd\@linechar
  \else\advance\x@diff \wd\@linechar\fi\fi
\ifnum\@yarg=0\relax\else\ifdim\y@diff >\z@\advance\y@diff -\ht\@linechar
  \else\advance\y@diff \ht\@linechar\fi\fi
\ifdim\x@diff <\z@ \@x@diff=-\x@diff \else\@x@diff=\x@diff\fi
\ifdim\y@diff <\z@ \@y@diff=-\y@diff \else\@y@diff=\y@diff\fi
\num@segments=0 \num@segmentsi=0
\ifdim\wd\@linechar >1pt
 \num@segmentsi=\@x@diff \divide\num@segmentsi \wd\@linechar\fi
\ifdim\ht\@linechar >1pt
 \num@segments=\@y@diff \divide\num@segments \ht\@linechar\fi
\ifnum\num@segmentsi >\num@segments \num@segments=\num@segmentsi\fi
\advance\num@segments \@ne %to account for round-off error
\ifnum #1=0 \relax \else\ifnum #1 < -99
  \typeout{***drawline: reduction <= -100 percent implies blankness!***}
\else\num@segmentsi=#1 \advance\num@segmentsi by 100
     \multiply\num@segments \num@segmentsi
     \divide\num@segments by 100
\fi\fi
\divide\x@diff \num@segments
\divide\y@diff \num@segments
\advance\num@segments \@ne %for the last segment for which I subtracted
\@xdim=#2\unitlength \@ydim=#3\unitlength
\if@negarg \advance\@xdim -\wd\@linechar\fi
\ifnum\@yarg <0 \advance\@ydim -\ht\@linechar\fi

\@killglue
\loop \ifnum\num@segments > 0
      \unskip\raise\@ydim\hbox to\z@{\hskip\@xdim \copy\@linechar\hss}%
      \advance\num@segments \m@ne\advance\@xdim\x@diff\advance\@ydim\y@diff%
\repeat \ignorespaces
\fi%the if of @toosmall
\fi}}% for \if@drawit

\newcounter{@sc} \newcounter{@scp} \newcounter{@t} \newlength{\@x}
\newlength{\@xa} \newlength{\@xb}  \newlength{\@y} \newlength{\@ya}
\newlength{\@yb} \newsavebox{\@pt}
\def\bezier#1(#2,#3)(#4,#5)(#6,#7){\c@@sc#1\relax
 \c@@scp\c@@sc \advance\c@@scp\@ne
 \@xb #4\unitlength \advance\@xb -#2\unitlength \multiply\@xb \tw@
 \@xa #6\unitlength \advance\@xa -#2\unitlength
 \advance\@xa -\@xb \divide\@xa\c@@sc
 \@yb #5\unitlength \advance\@yb -#3\unitlength \multiply\@yb \tw@
 \@ya #7\unitlength \advance\@ya -#3\unitlength
 \advance\@ya -\@yb \divide\@ya\c@@sc
 \setbox\@pt\hbox{\vrule height\@halfwidth depth\@halfwidth
 width\@wholewidth}\c@@t\z@
 \put(#2,#3){\@whilenum{\c@@t<\c@@scp}\do
 {\@x\c@@t\@xa \advance\@x\@xb \divide\@x\c@@sc \multiply\@x\c@@t
 \@y\c@@t\@ya \advance\@y\@yb \divide\@y\c@@sc \multiply\@y\c@@t
 \raise \@y \hbox to \z@{\hskip \@x\unhcopy\@pt\hss}\advance\c@@t\@ne}}}
\makeatother
\begin{document}

\def\BV{{\bf BV}}

\title{Generalized phase transitions in finite coupled map lattices}
\author{Michael Blank
       \\ \\
        Russian Ac. of Sci., Inst. for
        Information Transmission Problems, \\
        Ermolovoy Str. 19, 101447, Moscow, Russia, blank@obs-nice.fr}
\date{20.09.1995}
\maketitle

\n {\bf Abstract} -- We investigate generalized phase transitions of type
localization - delocalization from one to several Sinai-Bowen-Ruelle
invariant measures in finite networks of chaotic elements (coupled map
lattices) with general graphs of connections in the limit of weak
coupling. \bigskip

\section{Introduction} \label{cml.1}

During last decade a new class of models, so called networks of
chaotic elements or coupled map lattices (CML), has been introduced
to investigate complex dynamical phenomena in spatially extended
systems. A review of various applications of CML one can find in
\cite{Ka}. Mathematical investigation of ergodic properties of such
systems was started using methods and ideas of statistical physics in
\cite{BS2} and then was continued in \cite{PS,Bl17,KeK,BrK,Bl18}. The
main result obtained in these papers was the stability of statistical
properties of CML in the limit of weak coupling. In this paper we
discuss ergodic properties of finite CML with general graphs of
connections among elements in the limit of weak coupling and show
that in the general case there may be generalized phase transitions
in such systems of type localization - delocalization. These phase
transitions correspond to a situation, when trajectories, which
should normally be dense, remain confined to a small region (which
vanishes, when the coupling constant goes to zero). We call this
``localization phenomenon''. The first observation of this type was
published in \cite{Bl17} (see also \cite{KeK2} for numerical studies
of other phase transitions in the simplest dyadic models of CML). We
shall give several general statements about Sinai-Bowen-Ruelle (SBR)
measures of such systems and then we shall discuss in detail two
particular families of CML.

It is worthwhile to discuss the nature of the localization phenomenon.
Suppose for a moment that our uncoupled system is a hyperbolic map.
Consider a periodic trajectory of a map, in a small neighborhood of which
the map is strictly hyperbolic. Then near any point of this trajectory
local stable and unstable manifolds are going arbitrary close one to
another. Therefore stable and unstable directions can be mixed by means of
arbitrary small perturbations, and the result depends on whether the
contraction is stronger than expansion (localization), or not. CMLs that
we consider are not hyperbolic, but the same result gives the absence
of the local expanding property. This property means that a map expands
distances between any close enough points. This property distinguishes
the situations in Theorems~\ref{stab1:cml} and \ref{stab2:cml}. To show
that this phenomenon is not something obscure, specific for only
discontinuous maps, we shall prove its presence for the well known family
of quadratic maps in Section~\ref{square}.

Let $\{f_i\}$ be a sequence of one-dimensional nonsingular mappings
$f_i: X \to X=[0,1]$ from the unit interval into itself, let $X^d$ be
a direct product of these intervals, and let us denote a point in
this direct product by $\bar x = (x_1, \dots, x_d) \in X^d$. Remark
that the dimension $d$ needs not to be finite.

\begin{definition} By a {\em coupled map lattice} (CML) we shall mean
a map $F_\ep: X^d \to X^d$ defined as follows:
\be F_\ep(\bar x) = \Phi_\ep \circ F(\bar x), \ee
where the map $F$ is a direct product of the maps $f_i$ ( i.e.
$(F(\bar x))_i := f_i(x_i)$), and the map $\Phi_\ep: X^d \to X^d$, describing
the coupling, is defined as follows:
\be (\Phi_\ep \bar x)_i = (1-\ep) x_i + \ep \sum_j \gamma_{ij} x_j,
\label{coupling} \ee
and the matrix $\Gamma=(\gamma_{ij})$ is a stochastic matrix, describing
the graph of interactions among the maps. \end{definition}

Remark. The matrix $\Gamma$ needs not to be symmetric, which means that
the connections (couplings) may be oriented. We shall consider only
connected graphs, however there may be ``free'' nodes. We shall call
a node $i$ of the graph a free node, if $\gamma_{ij}=0$ for any $j$
(this does not contradict to the connectivity of the graph $\Gamma$).

The simplest known type of chaotic maps is so called piecewise
expanding (PE) maps.

%%%%%%%%%%%%%%%%%%%%%%%%%%%%%%%%%%%%%%%%%%%%%%%%%%%%%%%%%%%%%%%%%%
%% Picture for a ``typical'' PE map
\Bfig(150,150){\drawline(0,0)(150,0)(150,150)(0,150)(0,0)
               \drawline[-40](0,0)(150,150)
               \put(73,-10){$c$} \drawline[-40](75,2)(75,150)
               \put(113,75){$\la_f$}
               \put(0,-10){$0$} \put(150,-10){$1$} \put(-10,145){$1$}
               \thicklines \drawline(75,125)(150,20)
               \bezier{200}(0,0)(30,90)(75,140)
               %\thinlines \drawline(30,75)(70,110)
                }{A ``typical'' PE map. \label{gen.tent}}
%%%%%%%%%%%%%%%%%%%%%%%%%%%%%%%%%%%%%%%%%%%%%%%%%%%%%%%%%%%%%%%%%%

\begin{definition} We shall say that a map $f$ is {\em piecewise expanding}
(PE) if there exists a partition of the interval $X$ into disjoint
intervals $\{X_j\}$, such that $f\I|{\txt{Clos}(X_j)}$ is a
$C^2$-diffeomorphism (from the closed interval $\txt{Clos}(X_j)$ to its
image), and the {\em expanding constant} of the map
$$ \la_f := \inf_{j, \, x \in X_j} |f'(x)|, $$
is positive and for some integer $\kappa$ the iteration $f^\kappa$ of the
map $f$ has the expanding constant $\la_{f^\kappa}$ strictly larger than $1$.
\end{definition}

A typical example of a PE map is shown on Figure~\ref{gen.tent}. So
the map even needs not to be continuous. As it is well known,
starting from the paper of Lasota \& Yorke \cite{LY}, these maps have
all the statistical properties that one can demand on deterministic
dynamical system. It has smooth (absolutely continuous) invariant
measure $\mu_f$ (Sinai-Bowen-Ruelle measure of this map), exponential
correlation decay and CLT with respect to this measure, etc. In this
paper we want to discuss stability of these properties with respect to
small perturbations due to the coupling. We shall emphasize some aspects
of this problem, because they seem quite counterintuitive, at least from
our point of view.

\begin{definition} An {\em image} of a measure $\mu$ under the action of
a map $f$ is a new measure $f\mu$, such that $f\mu(A)=\mu(f^{-1}A)$ for any
measurable set $A$. \end{definition}

\begin{definition} Let there exists an open set $U$ in the phase space such
that for any smooth measure $\mu$ with the support in this set its images
$f^n \mu$ converge weakly to a measure $\mu_f$, do not depending on the
choice of the initial measure $\mu$. Then the measure $\mu_f$ is called
a {\em Sinai-Bowen-Ruelle (SBR)} measure of the map $f$. \end{definition}

\begin{definition} By a {\em singular point} of a map $f$ we shall mean
a point, where the derivative of the map is not well defined. The set
of singular points we shall denote by $\txt{Sing}[f]$.
(On Figure~\ref{gen.tent} the point $c$ is a singular point.)
\end{definition}

\n Typical examples of the graphs $\Gamma$ are shown on Figure~\ref{graphs}:
 - linear chain, cyclic chain, Cayley tree.

%%%%%%%%%%%%%%%%%%%%%%%%%%%%%%%%%%%%%%%%%%%%%%%%%%%%%%%%%%%%%%%%%%
%% Picture for ``typical'' graphs $\Gamma$
\Bfig(160,150){\drawline(0,0)(160,0)(160,150)(0,150)(0,0)
               \thicklines
               \put(15,130){\circle{5}} \put(55,130){\circle{5}}
               \put(95,130){\circle{5}} \put(135,130){\circle{5}}
               \drawline(15,130)(135,130) \put(145,130){(a)}
               {}
               \put(55,110){\circle{5}} \put(95,110){\circle{5}}
               \put(55,80){\circle{5}}  \put(95,80){\circle{5}}
               \put(30,95){\circle{5}}  \put(120,95){\circle{5}}
               \drawline(55,110)(95,110)(120,95)(95,80)(55,80)(30,95)(55,110)
               \put(145,90){(b)}
               {}
               \put(55,60){\circle{5}}  \put(95,60){\circle{5}}
               \put(75,50){\circle{5}}  \put(75,35){\circle{5}}
               \put(55,25){\circle{5}}  \put(95,25){\circle{5}}
               \put(35,25){\circle{5}}  \put(115,25){\circle{5}}
               \put(60,5){\circle{5}}   \put(90,5){\circle{5}}
               \drawline(75,35)(75,50)(55,60) \drawline(75,50)(95,60)
               \drawline(35,25)(55,25)(75,35)(95,25)(115,25)
               \drawline(55,25)(60,5)  \drawline(95,25)(90,5)
               \put(145,30){(c)}
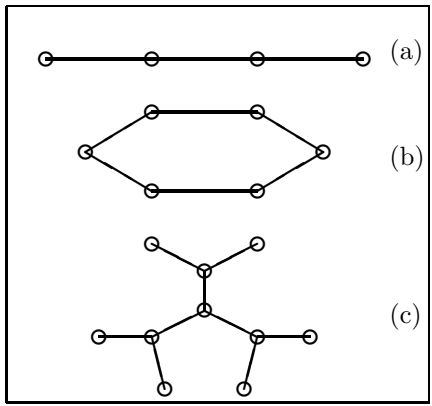
              }{``Typical'' graphs $\Gamma$: (a) linear chain, (b) cyclic
chain,
                (c) Cayley tree. \label{graphs}}
%%%%%%%%%%%%%%%%%%%%%%%%%%%%%%%%%%%%%%%%%%%%%%%%%%%%%%%%%%%%%%%%%%

Let us fix the sequence of maps $f_i$ and the matrix $\Gamma$. Clearly
for $\ep>0$ small enough the map $F_\ep$ is a piecewise expanding
$d$-dimensional map. The only difference between the multidimensional
case and the 1-dimensional one is that one should use the smallest
eigenvalue of Jacobi matrix $F'$ of a map $F$ as the expanding
constant of the map (see \cite{Bl5} for details). If we assume also
that $\la_{f_i} > \la$, which is large enough (actually much larger than
$1$), then all the ergodic properties of the corresponding CML follow
from general theory of multidimensional PE maps \cite{Bl5}. However, we
know only that the expanding constants are positive and larger than $1$
for some iteration of the map. Therefore general statements could not be
applied here and one should use the specific structure of the map $F_\ep$.
In the sequel we shall suppose that each map $f_i$ has only one SBR
measure $\mu_i$.

\begin{theorem} \cite{Bl17,KeK} \label{stab1:cml} Let $\la_{f_i} \ge
\la > 2$ for any integer $i$. Then for any $\ep$ small enough CML
$F_\ep$ has a smooth invariant SBR measure $\mu_\ep$ converging
weakly as $\ep \to 0$ to the direct product of one-dimensional smooth
SBR measures $\mu_i$. \end{theorem}

\begin{corollary} Let the assumptions of Theorem~\ref{stab1:cml} are valid.
Then for $\ep$ small enough CML $F_\ep$ also has the only one SBR measure.
\end{corollary}

The main results of this paper are the following statements, generalizing
Theorem~\ref{stab1:cml} and showing that at least some additional assumptions
are necessary for the stability of the direct product of SBR measures $\mu_i$.

\begin{theorem} \label{stab2:cml} Let for any $i$ the expanding
constants of PE maps $f_i$ are strictly positive $\la_{f_i}>\la>0$,
and there are no periodic singular points. Then for any $\ep>0$ small
enough the CML $F_\ep$ has a smooth invariant SBR measure $\mu_\ep$
converging weakly as $\ep \to 0$ to the direct product of
one-dimensional smooth SBR measures $\mu_i$. \end{theorem}

So the only topological obstacle for the stability of statistical
properties of CML with respect to weak coupling is the existence
of periodic singular points.

\begin{theorem} \label{instab1:cml} There exists a sequence of PE maps
$f_i$ with $\la_{f_i}>1$ and with periodic singular points such that for any
$\ep>0$ small enough CML $F_\ep$ has several smooth invariant SBR measures,
converging to singular invariant measures of the map $F$ as $\ep \to 0$.
\end{theorem}

\begin{remark} Under the assumptions of Theorem~\ref{instab1:cml} for
any $\ep>0$ small enough Lebesgue measure of the support of every ergodic
SBR measure of CML $F_\ep$ is of order $\ep^d$. \end{remark}

This theorem not only states the localization of invariant measures
(supports on small sets), but also guarantees their smoothness and
the absence of other SBR measures. Actually the localization
phenomenon was firstly shown in \cite{Bl17,Bl18}, where it was proved
that it is possible to construct CML such that its SBR measure is
localized in a small neighborhood of a fixed point, but the statement
about the smoothness of invariant measures, absence of other SBR
measures, and investigation of their properties are new. It is
worthwhile to remark that we assumed in Theorem~\ref{instab1:cml},
that all the expanding constants $\la_{f_i}$ are greater than one,
which is not the general case for CML, constructed of Lasota-Yorke
type maps. Therefore even more strong localization may take place,
when there are no smooth SBR measures for CML for any weak enough coupling.

\begin{theorem} \label{instab2:cml} There exists a sequence of PE maps
$f_i$ with periodic singular points such that for any $\ep>0$ small
enough CML $F_\ep$ has only singular invariant SBR measures,
converging to singular invariant measures of the map $F$ as $\ep \to 0$.
\end{theorem}

This statement shows that a system of chaotic maps can be stabilized
by arbitrary weak coupling. We shall show further that the opposite
statement is also true, i.e. a system of stable maps can become a
chaotic CML under arbitrary weak coupling.

These results may be considered as some kind of generalized phase
transitions, because when the coupling strength $\ep$ goes to zero we
can have two quite different types of behavior - there are two
different ``phases'' - with one SBR measure and with several of them.

A particular case of Theorem~\ref{stab2:cml} for cyclic graph
$\Gamma$ and identical maps $f_i$ with expanding constants strictly
greater than $1$ was proved earlier in a different way in \cite{Ku}.

\section{Functions of generalized bounded variation} \label{cml.2}

We mainly follow here \cite{Bl5,Bl17,Bl18}. Let $X^d=[0,1]^d \subset
\IR^d$ (or $d$-dimensional unit torus) with the uniform norm $|x| =
\max \{|x_i|: \, x \in \IR^d, \, i=1,\dots ,d\}$ and $d$-dimensional
Lebesgue measure $m=m_d$. Let us introduce the following notation:

$$ \mu(Y,h) = \int_Y h(x) d\mu(x); \quad \mu(Y) = \mu(Y,\1{}); \CR
   \Osc(h,Y,x) = \sup\{|h(x)-h(y)|: y \in Y\}; \CR
   \Osc(h,Y)=\sup\{\Osc(h,Y,x): x\in Y\}; \CR
   W(h,Y,t) = m(Y, \Osc(h, B_t(x) \cap Y, x)); \CR
   B_t(x) = \{y \in X^d: |x-y| \le t\}; \CR
   \var(h,Y) = \frac{1}{d} \limsup_{t\to 0}
     \Lpar \frac{1}{t} \inf_{\hat h \in h \subset \IL}W(\hat h,Y,t)\Rpar \CR
   \var(h) = \var(h,X^d), \quad ||h||_v = \var(h) + ||h||, \quad
   ||h|| = m(X^d,|h|). $$

\begin{definition} The functional $\var(h,Y)$ we shall call the generalized
{\em variation} (or simply variation) of the function $h$ over the set $Y
\subseteq X^d$ and the functional $Osc(h,Y)$ shall call {\em oscillation}. The
Banach space of integrable functions with bounded variation
$$ \{ h: X^d \to \IR^1 \, : \, \var(h) < \infty \}, $$
equipped with the variational norm
$$ ||h||_v = \var(h) + ||h|| $$
we shall denote by $\BV(X)$ (or simply by $\BV$) and shall call the
{\em space of functions of bounded variation}. \end{definition}

The following statement provides some elementary facts about the
the properties of these functionals.

\begin{lemma} \label{cml.Lemma 2.1.} Let functions $h, \, h_1, \, h_2$ be of
bounded variation and let $Y, \, Y_1, \, Y_2$ be connected closed subsets of
$X^d$. Then the following inequalities are valid:
\begin{enumerate}
\item $\var(h_1+h_2, Y) \le \var(h_1, Y) + var(h_2, Y)$.
\item $\var(h, Y_1 \cup Y_2) \le \var(h, Y_1) + \var(h, Y_2)$
      and this inequality becomes an equality if \ \
      ${\rm Clos}(Y_1) \cap {\rm Clos}(Y_2) = \emptyset$.
\item $\var(ch,Y) = c\,\var(h,Y)$ for any nonnegative number $c$.
\item $\var(h_1 \times h_2, Y) \le \var(h_1, Y)\,||h_2||_{\infty }
          + \var(h_2, Y)\,||h_1||_{\infty }$.
\item $\Osc(h,Y) \le \var(h,Y)$.
\item for any $x \in Y$
           $$ |h(x)| \le \var(h,Y) + \frac{1}{m(Y)} \int_Y |h(y)| \, dy). $$
\item let $Y = [a,b]$ then
    \be |h(a) + h(b)| \le \var(h,Y) + \frac{2}{m(Y)} \int_Y |h(x)| \, dx.
    \label{trace1} \ee
\item for any numbers $0 \le a_1 < a_2 < \dots < a_n \le 1$
    $$ \inf_{\hat h \in h \subset \IL} \sum_i |\hat h(a_i) - \hat h(a_{i+1}|
       \le \var(h, \, [a_1,a_n]). $$
\end{enumerate} \end{lemma}

\proof All the properties here are quite the same as for usual
one-dimensional variation. However to show how this technics works, let us
prove the only nontrivial items (5) - (8). Suppose that the set $Y$
is a segment. Let us fix $0 < t \ll 1$ and choose an integer $k>0$ such that
$$ (k-1)t < m(Y) \le  kt . $$
Then
$$ t \ge  m(Y)/k > t \times m(Y)/(t+m(Y)) . $$
Consider a partition of the segment $Y$ to consecutive disjoint open
intervals $Y_i$ of length $m(Y)/k$, such that the union of their closures
contains $Y$. Then
$$ W(h, Y,t) = m(Y, \Osc(h, B_t(x) \cap Y, x))
   \ge  m(Y, \sum_i \Osc(h, Y_i, x) \, \1{Y_i}(x)) $$
$$ = \sum_i m(Y_i)\,\Osc(h, Y_i) = \frac{m(Y)}{k}\,\Osc(h,Y) $$
$$ > t \frac{m(Y)}{t+m(Y)} \, \Osc(h,Y) $$
Hence for every $0 < t << 1$
$$ \Osc(h,Y) < \frac{t+m(Y)}{m(Y)} \frac{1}{t} W(h, Y,t) $$
Now going to the limit as $t\to 0$ we obtain the required
inequality for the case of the connected $Y$. The general case,
when the set $Y$ consists of several connected components is
reduced to the considered one by the restriction of the function
$h$ to each of the connected components of $Y$. The item (5) is proved.

The property (6) is a simple consequence of (5), because clearly the
value of the function could be estimated by the sum of its oscillation and
its means value.

To prove the item (7) set $c=(a+b)/2, \, e=(b-a)/2, \, E=[0,e]$ and
consider a function $H:E\to R^1$, defined via
$$ H(x) = |h(c-x)| + |h(c+x)|, \quad x \in E . $$
By the item (5)
$$ \Osc(H,E) \le \var(H,E) \CR
   \le \var(h, [a,c]) + \var(h, [c,b]) \le \var(h,Y) . $$
Besides, by the definition of the oscillation we have for all $x \in E$
$$ H(x) \le \Osc(H,E) + \frac{2}{b-a} \int_0^e H(x) \, dx \CR
   \le \var(h,Y) + \frac{2}{m(Y)} \int_Y |h(x)| \, dx , $$
because the second addend is equal to the mean value of the
function $|h|$. Thus, setting $x=e$ and using that $H(e) = |h(a) + h(b)|$,
we obtain the required inequality. The item (7) is proved.

It remains to prove the item (8).
$$ \sum_i |\hat h(a_i) - \hat h(a_{i+1}|
   \le \sum_i \Osc(h, \, [a_i,a_{i+1}]) $$ $$
   \le \sum_i \var(h, \, [a_i,a_{i+1}]) \le \var(h, \, [a_1,a_n]) $$
\qed

Remark, that the property (8) shows the correspondence between our
generalized variation and the usual one-dimensional variation.

We shall need also a multidimensional analogue of the
inequality~\eqrefB{trace1}.

\begin{lemma} \cite{Bl17,Bl18} \label{trace.lemma} Let $I^d \subseteq X$ be a
direct product of $d$ intervals: $I^d = I_1 \times \dots \times I_d$ (i.e.
$I^d$ is a rectangle). Then
\be \var(h \, \1{I^d}) \le 2\var(h,I^d)
    + \frac{2}{\min_j|I_j|} \int_X |h| \, dm_d \label{trace2} \ee
for any function $h: X \to \IR^1$ of bounded variation.
\end{lemma}

This statement follows from the estimate of the trace of a function on
a boundary $\delta I^d$ of the rectangle $I^d$:
\be \int_{\delta I^d} |h| \, dm_{d-1} \le \var(h,I^d)
    + \frac{2}{\min_j|I_j|} \int_X |h| \, dm_d \label{trace3} , \ee

\n where $m_{d-1}$ is $(d-1)$-dimensional Lebesgue measure on the surface
$\delta I^d$. It is worthwhile to remark that estimates of this type can be
obtained for arbitrary domains $Y \subseteq X$:
$$ \var(h \, \1{Y}) \le A(Y) \var(h,Y) + B(Y) \int_X |h| \, dm_d , $$
but the coefficients $A(Y)$ and $B(Y)$ depend crucially on the form of the
domain and can be arbitrary large even for small domains (actually they
even need not to be finite in the general case).

\section{Operator approach for CML. Generic case.}

Operator approach for CML is based on the investigation of the
Perron-Frobenius operator (transfer-matrix) $\IP_{F_\ep}$ describing
the action of the CML $F_\ep$ on densities of smooth measures on $X^d$.
For $\ep \ge 0$ small enough this operator leaves the space of functions
of bounded variation invariant, and the main idea, which was firstly
proposed in \cite{LY} for the investigation of one-dimensional PE maps,
is to obtain for arbitrary integers $k$ estimates of the following type:

\be \var(\IP_{F_\ep}^k h) \le C \alpha^k \var(h) + \beta ||h||
    \label{var.ineq} , \ee

\n where $0 < \alpha < 1$ and $C,\,\beta < \infty$. Remind, that for the
uncoupled map $F$ ($\ep=0$), the analogous inequality:

\be \var(\IP_F^k h) \le C_0 \alpha_0^k \var(h) + \beta_0 ||h||
    \label{var.ineq0} , \ee

\n immediately follows from the well known properties of the
Perron-Frobenius operators for the one-dimensional PE maps $f_i$. Now
if the operator $\IP_{F_\ep}$ satisfies the
inequality~\eqrefB{var.ineq}, then a standard technics, based on
abstract ergodic theorem due to Ionescu-Tulchea and Marinescu (see
the suitable for this approach variant in \cite{Bl5}) gives a
possibility to obtain the spectral decomposition of this operator as
a sum of a contraction and a finite dimensional projector and to
obtain all the standard statistical properties.

\begin{remark} The class of coupling maps $\Phi_\ep$ should be defined up to
a nonsingular (piecewise smooth) conjugation. \end{remark}

Indeed, suppose that $\tilde f := \phi f \phi^{-1}, \; y=\phi x$ and
$x \to \Phi_\ep \circ F x$ then
$$ y \to (\phi \circ \Phi_\ep \circ \phi^{-1}) \circ \tilde F, $$
where by $\tilde F$ we mean a direct product of the conjugated maps
$\tilde f_i$:
$$ \tilde F := \phi \circ F \circ \phi^{-1} .$$

Let us start from the situation without the periodic singular points. Consider
the Perron-Frobenius operator, corresponding to CML:
$$ \IP_{F_\ep} = \IQ_\ep \IP_F ,$$
where the operator $\IQ_\ep$ corresponds to the coupling, and $\IP_F$ is
the Perron-Frobenius operator for the uncoupled system. Suppose for a moment
that these two operators are commutative. Then
$$ \IP_{F_\ep}^k = \IQ_\ep^k \IP_F^k . $$

\n From the definition of CML it follows that the coupling operator
$\IQ_\ep$ is close to identity in the following strong sense:

\be ||\IQ_\ep||_v \le 1+B\ep , \label{identity} \ee

\n where the constant $B < \infty$ does not depend on the parameter
$\ep>0$ for $\ep$ small enough. Therefore

\be ||\IP_{F_\ep}^k||_v \le ||\IQ_\ep^k h||_v ||\IP_F^k h||_v
    \le C (1+B\ep)^k \alpha_0^k ||h||_v + (1+B\ep)^k \beta_0 ||h|| .
\label{pert.commut} \ee

\n And thus, choosing $k$ large enough, such that
$$ \alpha:=C (1+B\ep)^k \alpha_0^k < 1, $$
we shall have the main inequality.

Unfortunately, these two operators are not commutative in the general case.
Therefore to apply this idea we shall construct a new operator
$\tilde \IQ_\ep$, such that
$$ \IP_{\tilde F} \tilde \IQ_\ep = \IQ_\ep \IP_F , $$
to use the operator $\tilde \IQ_\ep$ instead of $\IQ_\ep$.

Consider a partition of the phase space $X^d$ into rectangles $\Delta_j$,
such that the restriction of the map $F$ to $\Delta_j$ is a diffeomorphism
(i.e. every $\Delta_j$ is a direct product of some intervals of monotonicity
for the maps $f_i$). Then

\be \IP_F h = \sum_j \IP_{F,j} h \label{sum.monot} ,\ee

\n where

\be \IP_{F,j} h(x) := h(F_j^{-1}x) \, |\det (F_j^{-1}x)'| \, \1{F \Delta_j}(x),
\label{local.operator} \ee

\n $F_j:=F\I|{\Delta_j}$, and $(Fx)'$ is the Jacobi matrix of the map $F$.

Then on a small rectangular extension $\tilde \Delta_j$ of the rectangle
$\Delta_j$  the operator $\tilde \IQ_\ep$ is defined as follows:

\be \tilde \IQ_\ep h_{\tilde \Delta_j}
   = \IP_{\tilde F_j^{-1}}(\IQ_\ep \IP_F h_{\Delta_j}), \ee

\n where $h_{\Delta_j}$ is the restriction of a function $h$ to the set
$\Delta_j$, and $\tilde F_j$ is a $C^2$-smooth extension to a
diffeomorphism of the map $F_j$ on $\tilde \Delta_j$. The collection of
maps $\tilde F_j$ on the overlapping rectangles $\tilde \Delta_j$ may be
considered as a multivalued map $\tilde F$ from the unit cube into
itself. Denote the restriction of the operator $\tilde \IQ_\ep$ to the
$j$-th branch of the map by $\tilde \IQ_{\ep,j}$, and let us define a
function $J(x):=|\det (\tilde F_j^{-1}x)'|$. Then

$$ \tilde \IQ_{\ep,j} h(x) = J(x) \IQ_\ep \IP_F h(\tilde F x) $$

If we prove now that the operator $\tilde \IQ_\ep$ is also close to
identity (i.e. we prove an inequality of type (\ref{identity}) with
may be another constant, say $B'$, which also will not depend on $\ep$),
then the argument above can be applied in this case. To prove this
inequality we mainly follow the idea introduced in our paper \cite{BK}
in the investigation of random perturbations of one-dimensional PE maps.
Therefore here we shall discuss in detail only the first step of the
estimation, where the difference between the multidimensional case
(considered in this paper) and one-dimensional (considered in \cite{BK})
really takes place.

Let us fix some integer $j$ and consider only the branch $F_j$ on
the rectangle $\Delta_j$. For the sake of simplicity we change the
notation here and drop the index $j$, i.e. we shall denote $F_j$ by $F$
and $\Delta_j$ by $\Delta$. Let for some point $y \in \Delta$ we
have
$$ \var\left(\frac{J}{J(y)}\right),\ \var\left(\frac{J(y)}{J}\right)
\le\beta\ , $$
and let $\BV_0:=\{h \in \BV(X), \; h(u)=0: \; u\in \delta X^d \}$.
Then $||h\||_\infty \le \var(h)$ for all $h \in \BV_0$.

\begin{lemma} Let $\IQ$ be a (sub)Markov operator satisfying
$$ \var(\IQ h) \le \var(h) + C \cdot ||h|| $$
for all $h\in\BV_0$. Let
$$ \tilde \IQ:=\IP_{\tilde F_i^{-1}} \IQ \IP_{F_i}, $$
then $\IP_{\tilde F} \tilde \IQ = \IQ \IP_{F}$ and
$$ \var(\tilde \IQ h) \le (1+\beta)^2 \cdot \var(h, \Delta)
    +C(1+\beta) ||J||_\infty \cdot ||h \cdot \1{\Delta} || $$
for all $h\in\BV_0$. \end{lemma}

\proof Let $g(x):=\frac{J(x)}{J(y)}$. Then
$\var(g) \le \beta$, \, $||g||_\infty \le 1+\frac \beta 2$ and
$$ \tilde Qh(x) = g(x) \cdot \IQ \IP_F h(J(y) \cdot h) (\tilde F x) $$
because $\IQ \IP_F$ is a linear operator. Therefore
$$ \var(\tilde \IQ h)
     \le \var(g) \cdot ||\IQ \IP_F(J(y) \cdot h)||_\infty
         + ||g||_\infty \cdot \var(\IQ \IP_F(J(y) \cdot h)) \CR
     \le \beta \cdot \frac 12 \var(\IQ \IP_F(J(y) \cdot h))
         +(1 + \frac\beta 2) \cdot \var(\IQ \IP_F(J(y) \cdot h)) \CR
     \le (1+\beta) \cdot \left(\var(\IP_F(J(y) \cdot h))
        +C \cdot ||\IP_F(J(y) \cdot h)|| \right) \CR
     \le (1+\beta) \cdot \left(\var((\frac{J(y)}{J} \cdot h)
         \circ F^{-1} \cdot \1{F \Delta})
         + C \cdot ||J(y) \cdot h \cdot \1{\Delta} || \right) \CR
     \le (1+\beta) \cdot \left(\var(\frac{J(y)}{J} \cdot h \cdot \1{\Delta})
         + C \cdot ||J||_\infty \cdot ||h \cdot \1{\Delta}|| \right) \CR
     \le (1+\beta) \cdot \left(\var(\frac{J(y)}{J})
         ||h \cdot \1{\Delta}||_\infty
         + |\frac{J(y)}{J}\|_\infty \var(h \cdot \1{\Delta})
         + C \cdot ||J||_\infty \cdot ||h \cdot \1{\Delta}|| \right) \CR
     \le (1+\beta) \cdot \left(\beta \cdot \frac 12 \var(h \cdot \1{\Delta})
         + (1+\frac\beta 2) \cdot \var(h \cdot \1{\Delta})
         + C \cdot ||J||_\infty \cdot ||h \cdot \1{\Delta}|| \right) \CR
     \le (1+\beta)^2 \var(h \cdot \1{\Delta})
         + (1+\beta) C ||J||_\infty \cdot ||h \cdot \1{\Delta}|| $$
\qed

\begin{remark} To obtain an arbitrary small value of $\beta$ above it
is enough to choose fine enough subpartitions for our maps $f_i$.
\end{remark}

This finishes the proof of Theorem~\ref{stab2:cml}.

It seems, that the same idea could be applied for a general
multidimensional piecewise expanding map. However, there are
two obstacles. The first one is pure topological - existence of
periodic singular points, which we shall discuss in the next section.
The second obstacle is due to the fact that in the general case
elements of the partition for $F$ (and especially for $F^n$) are not
rectangles. As we mentioned before this can lead to large (and even
infinite) coefficients in the corresponding estimates for the
Perron-Frobenius operator $\IP_F$.

\section{Operator approach for CML. Localization for the case with
periodic singular points.} \label{examples}

Consider now the case with periodic singular points. To prove
Theorems~\ref{instab1:cml} and \ref{instab2:cml} it is enough to
construct examples of CML with periodic singular points, satisfying
the assumptions of these theorems. We shall show that it is enough to
consider two-dimensional case ($d=2$) with identical maps
$f_1=f_2=f$, having only two fixed singular points $c_1,c_2$ and the
following symmetrical coupling:
$$ c_1=f(c_1) \ne f(c_2)=c_2, \quad \gamma_{11}=\gamma_{22}=0,
                              \quad \gamma_{12}=\gamma_{21}=1.$$

Consider the following two families of maps:

$$ f_{b,c}^{(1)}(x)=\cases{
                   \frac12 - \frac{x/2}{c-c/b}, &if $0 \le x \le c-c/b$; \cr
                   b(x-c)+c,            &if $c-c/b \le x \le c$; \cr
                  -b(x-c)+c,            &if $c \le x \le c+c/b$; \cr
                   \frac{x+c+c/b}{1-2c-2c/b}, &if $c+c/b \le x \le 1/2$; \cr
                   1 - f(1-x),          &if $1/2 \le x \le 1$. \cr} $$

$$ f_{b,c}^{(2)}(x)=\cases{
                  -\frac{b-c}{c}x + b,    &if $0 \le x \le c$;   \cr
                  -\frac{c}{0.5-c}x +0.5, &if $c \le x \le 1/2$; \cr
                   1 - f(1-x),            &if $1/2 \le x \le 1$. \cr} $$

$$ \la_{f_{b,c}^{(1)}}:=b>1, \quad
   \la_{f_{b,c}^{(2)}}:=
       \min \Lbrace \frac{c}{\frac{1}{2}-c}, \; \frac{b-c}{c} \Rbrace <1.$$

$$ \la_{(f_{b,c}^{(2)})^2}:=\frac{b-c}{\frac{1}{2}-c} > 1 $$

%\newpage
%\vskip 1cm

%%%%%%%%%%%%%%%%%%%%%%%%%%%%%%%%%%%%%%%%%%%%%%%%%%%%%%%%%%%%%%%%%%
%% Pictures for CML localization
\Bfig(350,150)
    {% picture (a)
     \drawline(0,0)(150,0)(150,150)(0,150)(0,0)
     \drawline[-50](0,0)(150,150) %\drawline[-50](200,0)(350,150)
     \drawline[-50](35,0)(35,35) \drawline[-50](115,115)(115,0)
       \put(33,-8){$c$} \put(50,-8){$c+c/b$}
       \put(107,-8){$1-c$}  \put(38,10){$\la$}
       \put(-8,70){$\frac{1}{2}$} \put(70,-20){{\bf(a)}}
     \thicklines
     \drawline(0,75)(15,0)(35,35)(55,0)(95,150)(115,115)(135,150)(150,75)
     % picture (b)
     \put(200,0){\bpic(150,150){\thinlines
     \drawline(0,0)(150,0)(150,150)(0,150)(0,0)
     \drawline[-50](0,0)(150,150)
     \drawline[-50](30,0)(30,30) \drawline[-50](120,120)(120,0)
                                 \drawline[-50](75,0)(75,150)
     \drawline[-25](0,100)(50,50)(75,0)
     \drawline[-25](75,150)(100,100)(150,50)
       \put(28,-8){$c$}  \put(110,-8){$1-c$} \put(73,-10){$1/2$}
       \put(-8,100){$b$} \put(152,50){$1-b$} \put(70,-20){{\bf(b)}}
     \thicklines
     \drawline(0,100)(30,30)(75,0) \drawline(75,150)(120,120)(150,50)
     }}
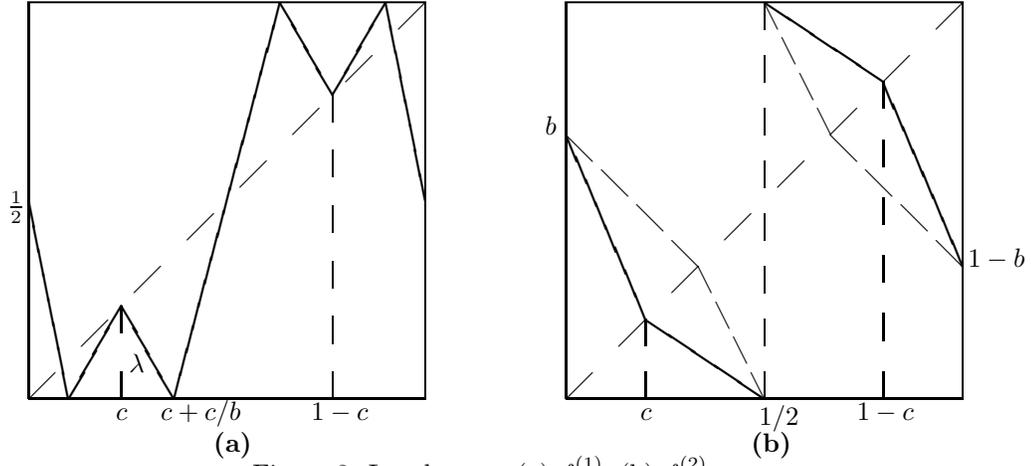
    }{Local maps, (a) $f_{b,c}^{(1)}$, (b) $f_{b,c}^{(2)}$ \label{loc.maps}}

%\vskip 2cm

\Bfig(350,150)
    {% picture (a)
     \drawline(0,0)(150,0)(150,150)(0,150)(0,0)
     \drawline[-50](35,0)(35,150) \drawline[-50](115,0)(115,150)
     \drawline[-50](0,35)(150,35) \drawline[-50](0,115)(150,115)
     \drawline[-10](25,0)(25,150) \drawline[-10](45,0)(45,150)
     \drawline[-10](0,25)(150,25) \drawline[-10](0,45)(150,45)
     \drawline[-10](105,0)(105,150) \drawline[-10](125,0)(125,150)
     \drawline[-10](0,105)(150,105) \drawline[-10](0,125)(150,125)
     \put(22,-8){$c_-$} \put(42,-8){$c_+$} \put(120,-8){$1-c_-$}
     \put(30,110){$K_1$} \put(110,30){$K_2$} \put(70,-20){{\bf(a)}}
     \thicklines
         \drawline(25,105)(45,105)(45,125)(25,125)(25,105)
         \drawline(105,25)(125,25)(125,45)(105,45)(105,25)
     \thinlines
     % picture (b)
     \put(200,0){\bpic(150,150){
     \drawline(0,0)(150,0)(150,150)(0,150)(0,0)
     \drawline[-50](30,0)(30,150) \drawline[-50](120,0)(120,150)
     \drawline[-50](0,30)(150,30) \drawline[-50](0,120)(150,120)
     \drawline[-10](40,0)(40,150) \drawline[-10](60,0)(60,150)
     \drawline[-10](0,40)(150,40) \drawline[-10](0,60)(150,60)
     \drawline[-10](90,0)(90,150) \drawline[-10](110,0)(110,150)
     \drawline[-10](0,110)(150,110) \drawline[-10](0,90)(150,90)
     \put(37,-8){$c_-$} \put(57,-8){$c_+$} \put(100,-8){$1-c_-$}
     \put(45,95){$K_1$} \put(95,45){$K_2$} \put(70,-20){{\bf(b)}}
     \thicklines
         \drawline(40,90)(60,90)(60,110)(40,110)(40,90)
         \drawline(90,40)(110,40)(110,60)(90,60)(90,40)
     \thinlines
     }}
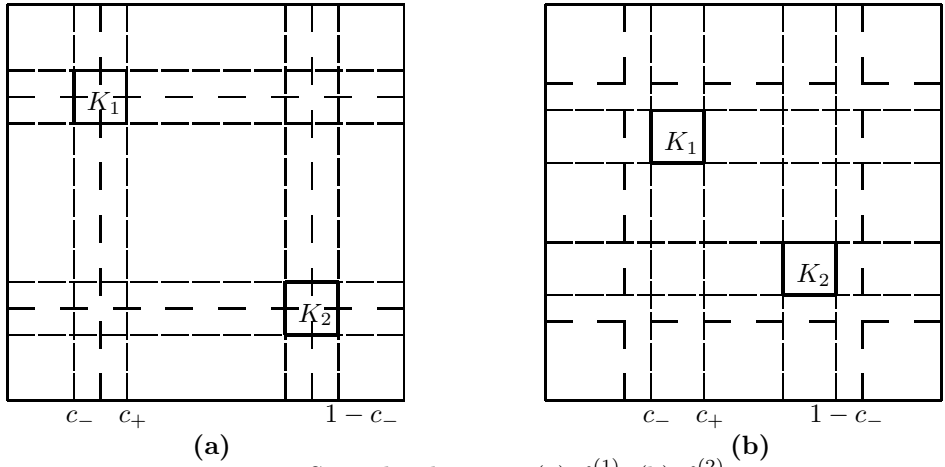
     }{Space localization: (a) $f_{b,c}^{(1)}$, (b) $f_{b,c}^{(2)}$
       \label{loc.maps.space}}

%\vskip 2cm

%%%%%%%%%%%%%%%%%%%%%%%%%%%%%%%%%%%%%%%%%%%%%%%%%%%%%%%%%%%%%%%%%%
%% (a) Schematic action of coupling near fixed points,
%% (b) bifurcation diagram for the map $f_{b,c}^{(2)}$
\Bfig(350,150)
    {% picture (a)
     \drawline(0,0)(150,0)(150,150)(0,150)(0,0)
     \drawline[-40](0,0)(150,150) \drawline[-20](75,0)(75,150)
     \thicklines \drawline(0,150)(40,0)(75,75)(110,0)(150,150)
     \drawline[-30](40,15)(75,90)(110,15)
     \put(73,-8){$c$} \put(55,20){$\la_f$} \put(70,-20){{\bf(a)}}
     \thinlines
     % picture (b)
     \put(200,0){\bpic(150,150){
     \drawline(0,0)(150,0)(150,150)(0,150)(0,0)
     \drawline(75,0)(75,150) \drawline(0,75)(150,75)
     \put(-4,-8){$0$} \put(70,-10){$\frac{1}{4}$}
     \put(145,-10){$\frac{1}{2}$}  \put(155,-4){$c$}
     \put(-8,75){$\frac{1}{2}$} \put(-8,145){$1$} \put(-4,155){$b$}
     \put(25,110){$C \to S$} \put(100,30){$S \to S$}
     \put(35,30){$S$}        \put(110,110){$C$}
     \put(70,-20){{\bf(b)}}
     \thinlines
     }}
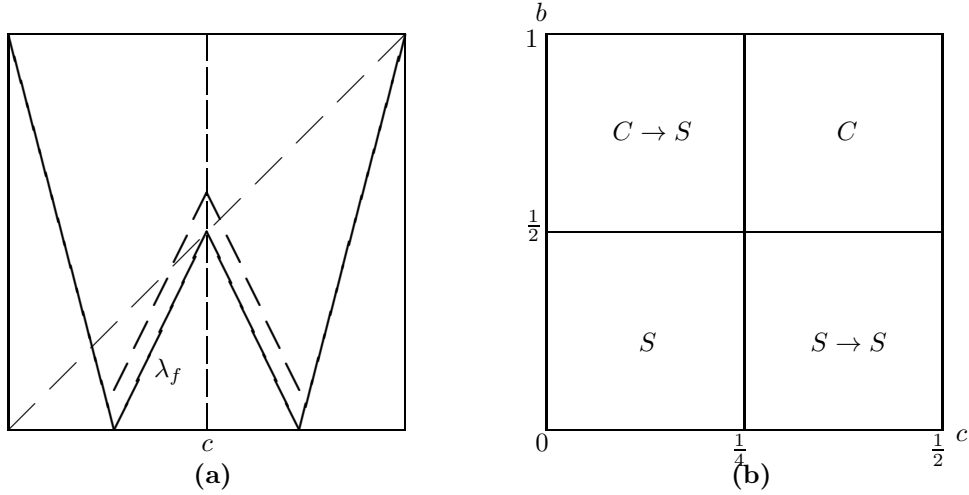
     }{(a) Local action of coupling near fixed singular points,
       (b) bifurcation diagram for the map $f_{b,c}^{(2)},
       \; 0 \le b \le 1, \; 0 \le c \le 1/2$ \label{bif.diag}}
%%%%%%%%%%%%%%%%%%%%%%%%%%%%%%%%%%%%%%%%%%%%%%%%%%%%%%%%%%%%%%%%%%%%%%

Let us start the investigation from the first family $f_{b,c}^{(1)}$.
Fix some $1<b<2, \, 0<c<1/2, \, \ep>0$. Then the expanding constant
$\la_f=b>1$. Consider the first two points of the trajectory of the
point $(c,1-c)$ of the CML $F_\ep$ constructed by means of this map:
$$ \Lpar c \above0pt {1-c} \Rpar \farrow{f,\Phi_\ep}
   \Lpar {c + \ep (1-2c)} \above0pt {1-c - \ep (1-2c)} \Rpar \farrow{f}
   \Lpar {c - \ep b (1-2c)} \above0pt {1-c - \ep b (1-2c)} \Rpar \CR
   \farrow{\Phi_\ep}
   \Lpar {c - \ep (b (1-2c)+2c-1) + 2\ep^2 b (1-2c)} \above0pt
         {1-c + \ep (b (1-2c)+2c-1) - 2\ep^2 b (1-2c)} \Rpar $$

Denote now
$$ c_-:=c - \ep (b-1)(1-2c), \quad  c_+:=c + \ep (1-2c) \CR
   K_1:=[c_-, c_+] \times [1-c_+, 1-c_-], \quad
   K_2:=[1-c_+, 1-c_-] \times [c_-, c_+] $$

\begin{lemma} \label{l_1} $F_\ep K_i \subset K_i$ for every $i=1,2$.
\end{lemma}

\proof The restriction of the map $F_\ep$ to the rectangle $K_1$ is a
continuous map, such that the boundary of the rectangle $K_1$
is mapped by $F_\ep$ into $K_1$ and at least one inner point of $K_1$ (the
point $(c,1-c)$) is also mapped by $F_\ep$ into $K_1$. To show this, it is
enough to prove that every corner of the rectangle $K_1$ is mapped into
$K_1$. Let us prove it for the left lower corner $(c_-, 1-c_+)$.
Denote $t:=\ep (1-2c), \; \beta:=b-1<1$. Then $c_-=c-t\beta, \;
c_+=c+t$ and
$$ \Lpar c_- \above0pt {1-c_+} \Rpar
   = \Lpar {c-t\beta} \above0pt {1-c-t} \Rpar
   \farrow{f} \Lpar {c - b t \beta} \above0pt {1-c + b t} \Rpar \CR
   \farrow{\Phi_\ep}
   \Lpar {c - b t \beta + \ep(1-2c+tb+tb\beta)} \above0pt
         {1-c+tb - \ep(1-2c+tb+tb\beta)} \Rpar \CR
   = \Lpar {c - tb + tb - b t \beta + \ep(1-2c+tb+tb\beta)} \above0pt
           {1-c -t + t + tb - \ep(1-2c+tb+tb\beta)} \Rpar \CR
   = \Lpar {c - tb + tb - b t \beta + \ep(1-2c+tb+tb\beta)} \above0pt
           {1-c-t + tb(1 - \ep(1+\beta))} \Rpar \CR
   > \Lpar {c - tb} \above0pt {1-c-t} \Rpar
   = \Lpar c_- \above0pt {1-c_+} \Rpar , $$
where by $\xi>\eta$ we mean that every coordinate of the vector $\xi$
is greater than the corresponding coordinate of the vector $\eta$.

On the other hand,
$$ \Lpar {c - b t \beta + \ep(1-2c+tb+tb\beta)} \above0pt
         {1-c+tb - \ep(1-2c+tb+tb\beta)} \Rpar \CR
   = \Lpar {c+t - b\ep((1-2c)\beta-t-t\beta)} \above0pt
         {1-c+t - \ep((2-b)(1-2c)+tb+tb\beta)} \Rpar \CR
   < \Lpar {c+t} \above0pt {1-c+t} \Rpar
   = \Lpar c_+ \above0pt {1-c_-} \Rpar . $$

\n In the same way one can show that three other corner points are
mapped into the rectangle $K_1$. The proof for the rectangle $K_2$ is
analogous. Therefore these rectangle are mapped into themselves under
the action of $F_\ep$. \qed

\begin{lemma} \label{l_2} The restriction of the map $F_\ep$ to a
rectangle $K_1$ ($K_2$) has a smooth invariant measure (SBR measure)
$\mu_1$ ($\mu_2$). \end{lemma}

\proof Denoting $c_1=c, \; c_2=1-c$ we consider the following change of
variables:
\be y_i:=\frac{x_i-c_i}{\ep}, \label{change} \ee
which gives new local maps $\tilde f_i$ from the neighborhoods of
the points $c_i$ (whose direct product gives the rectangle $K_1$) into
some new intervals with sides of order $1$:

\be y_i \to (1-\ep) \tilde f(y_i) + C_i + {\cal O}(\ep)
\label{renorm} ,\ee

\n where $\tilde f_i(0)=0$ and $\prod_i C_i \ne 0$. The shape of the new
local map $\tilde f$ in a neighborhood of a fixed singular point is shown
by a dotted line on Figure~\ref{bif.diag}.a. This new system is also a couple
map lattice, but without singular periodic points (for $\ep>0$ small
enough). Therefore we can apply here the results of Theorem~\ref{stab2:cml}
to obtain the existence of smooth invariant measures. \qed

\begin{lemma} \label{l_3} There are only two SBR measures $\mu_i, \; i=1,2$
of the map $F_\ep$ and their supports $\txt{supp}(\mu_i) \subset K_i$.
\end{lemma}

\newcommand\FF{{\tilde F}}

\proof Let us construct a new map $\FF$ from $X^2$ into
itself, which will differ from the map $F$ only on the rectangles $K_i$:
$$ \FF(x) = \cases{F(x)-2\ep(1,-1), &if $x \in K_1$; \cr
                   F(x)+2\ep(1,-1), &if $x \in K_2$; \cr
                   F(x),        &otherwise. \cr} $$
Then for $\ep>0$ small enough the map $\FF_\ep:=\Phi_\ep \circ \FF$ is a PE
map. The difference between this map and the map $F_\ep$ is that, it has
no ``traps'' around singular points $c$ and $1-c$. Now using the same
argument as in the proof of Theorem~\ref{stab2:cml} one can prove the
existence and the uniqueness of a smooth invariant measure $\mu_{\FF}$ of
this map, which will have a positive density on the whole rectangle
$X^2$. Therefore a.a. trajectory of the map $\FF_\ep$ is dense on $X^2$.

This shows that a.a. trajectory of the original map $F_\ep$ hits eventually
into one of the sets $K_i$, because it coincides with some trajectory
of the map $\FF_\ep$ up to the moment, when it hits into one of the
sets $K_i$. \qed

This finishes the proof of Theorem~\ref{instab1:cml}. \bigskip

In a close way one can investigate also the second family of
maps $f_{b,c}^{(2)}$. We shall use the same notation as for the
first family. Fix some $1/2<b<1, \, 0<c<1/2, \, \ep>0$. Then the expanding
constant $\la_{f^2}=(b-c)/(1/2-c)>1$, however
$\la_f=\min\{(b-c)/c, c/(1/2-c)\}$ may be less than 1. Consider the
first two points of the trajectory of the point $(c,1-c)$ of
the CML $F_\ep$ constructed by means of this map:
$$ \Lpar c \above0pt {1-c} \Rpar \farrow{f,\Phi_\ep}
   \Lpar {c + \ep (1-2c)} \above0pt {1-c - \ep (1-2c)} \Rpar \farrow{f}
   \Lpar {c - 2\ep c} \above0pt {1-c - 2\ep c} \Rpar \CR
   \farrow{\Phi_\ep}
   \Lpar {c - \ep (4c-1) + 4\ep^2 c} \above0pt
         {1-c + \ep (4c-1) - 4\ep^2 c} \Rpar $$
Denote now
$$ c_-:=c - \ep (4c-1), \quad  c_+:=c + \ep (1-2c) \CR
   K_1:=[c_-, c_+] \times [1-c_+, 1-c_-], \quad
   K_2:=[1-c_+, 1-c_-] \times [c_-, c_+] $$

\begin{lemma} \label{l_4} Let $4c<1$, then $F_\ep K_i \subset K_i, \; i=1,2$.
\end{lemma}

The proof of this statement is analogous to the proof of Lemma~\ref{l_1}.

\begin{lemma} \label{l_5} Let $4c<1$, then restriction of the map $F_\ep$ to a
rectangle $K_1$ ($K_2$) has a globally stable fixed point $p_1$ ($p_2$).
\end{lemma}

\proof The main difference between the considered situations and the result
of Lemma~\ref{l_2} is that the restriction of this CML to one of the
rectangles $K_i$ is not a PE map, because  for any its iterations the
expanding constant is strictly less than $1$. Indeed, the derivative of the
local map is equal to $c/(1/2-c)<1$ for $4c<1$, and the coupling can only
decrease the expansion. \qed

Theorem~\ref{instab2:cml} is proved. \bigskip

Using the same argument as in the proof of Lemmas~\ref{l_1},~\ref{l_2}, one
can show that the situation will be quite different when $0<b<1/2$. The
point is, in this region of parameters the uncoupled system is not
expanding and has two locally stable fixed points.

\begin{lemma} \label{l_6} Let $4c>1$ and $0<b<1/2$, then
$F_\ep K_i \subset K_i, \; i=1,2$ and CML $F_\ep$ has two singular SBR
measures with the supports on cycles of period 2.
\end{lemma}

The bifurcation picture for the second family of maps is shown on
Figure~\ref{bif.diag}.b. The uncoupled system is chaotic when $b>1/2$ and
stable otherwise. We denoted this two possibilities by letters ``C'' and
``S''. Consider the coupled system for a small enough $\ep>0$.
Lemmas~\ref{l_4},~\ref{l_5} show that when $4c<1$ there is a transition
from the chaotic behavior to the stable one ($C \to S$). Lemma~\ref{l_6}
describes another transition from stable fixed points to stable
period 2 cycles ($S \to S$). When $4c<1, \; 0<b<1/2$ or $4c>1, \; 1/2<b<1$
there are no ``traps'' around the fixed singular points and therefore there
are no ``phase'' transitions.

Generalization of our results to the general case of periodic
singular points (instead of fixed points) and arbitrary dimension $d$
is straightforward and it is easy to formulate some simple sufficient
conditions of the localization phenomenon. However any CML, whose
local map has only one fixed singular point, shows the absence of
the localization. Therefore the investigation of necessary conditions is
a more complex problem and will be published elsewhere.

%%%%%%%%%%%%%%%%%%%%%%%%%%%%%%%%%%%%%%%%%%%%%%%%%%%%%%%%%%%%%%%%%%%%%%

\section{Localization in CML, constructed by smooth maps} \label{square}

Actually the appearance of the localization in CML is quite unexpected and
may seem to be a consequence of the discontinuity of PE maps. To show that
it is not so and that this phenomenon is not artificial and is generic, we
shall prove in this section its existence for smooth local maps. The
simplest way to do it is to smooth the maps $f_{b,c}^{(i)}$ near periodic
singular points. Howether we shall consider a more general situation - a
well known family of quadratic maps $f_a(x):=ax(1-x), \; 3<a\le 4$. This
situation differs from the results of
Teorems~\ref{instab1:cml},\ref{instab2:cml} in that sense, that the
localization takes place only for nonzero coupling strength. Let the value
of the parameter $a$ is close enough to $4$ and the map $f_a$ has a smooth
SBR measure. It is well known that the set of such values of the parameter
$a$ is of positive Lebesgue measure. We shall consider a CML, constructed
by means of 2 identical square maps with the parameter $a$.

\begin{lemma} \label{ls1} There exists a constant $3<a_0\le4$, and an
inteval of values of the coupling strength $0.14 < \ep_1 < \ep < \ep_2 < 0.2$
such the the CML $F_\ep$ has a stable periodic trajectory with period $2$:
$$ F_\ep(p_1,p_2) = (p_2,p_1); \quad F_\ep(p_2,p_1) = (p_1,p_2), $$
where $p_1<1/2<p_2<1$ and $f'_a(p_1) \times |f'_a(p_2)| > 1$, and therefore
an SBR measure, localized on this trajectory. \end{lemma}

\proof We can follow the same construction as in the case of PE maps to
find two ``traps'' for CML, but in the case of square maps this way is too
complex. Therefore we shall use the fact that the map $f_a$ is smooth and
smoothly depends on the parameter $a$. Therefore it is enough to prove the
existence of a stable periodic trajectory for a given pair of values
$(a,\ep)$. Let us set $a=4, \; \ep=.17$. A simple calculation shows that
the pair of points $p_1=0.484989\dots$ and $p_2=0.893799\dots$ defines the
periodic trajectory with period $2$ for this pair of values $(a,\ep)$. To
prove its stability we shall calculate Jacobian matrices of the CML in this
poits and shall show that eigenvalues of their product are positive and
less than $1$. The Jacobian matrices and their product are:
$$ \Lpar {{0.0997 \;\; 0.5356} \above0pt {0.0204 \;\; 2.6148}} \Rpar \quad
   \Lpar {{2.6148 \;\; 0.0204} \above0pt {0.5356 \;\; 0.0997}} \Rpar \quad
   \Lpar {{0.5476 \;\; 0.0646} \above0pt {1.4538 \;\; 0.2904}} \Rpar $$
The eigenvalues of the product matrix are $0.7512$ and $0.0867$. \qed

%%%%%%%%%%%%%%%%%%%%%%%%%%%%%%%%%%%%%%%%%%%%%%%%%%%%%%%%%%%%%%%%%%%%%%

\section{Localization phenomenon for CML with a large number of elements}

Consider CML with a large number of elements. Suppose that all
the maps $f_i$ are identical and correspond to one of the two
families considered in Section~\ref{examples}, i.e. the map $f$ has
two fixed singular points $c$ and $1-c$. Consider a sequence of numbers
$\bar w = \{w_j\}$, where $0 \le w_j \le 1$. We shall assume that

\be  \Omega(\bar w,i) := \sum_j \gamma_{ij} w_j - w_i \label{cond.top} \ee

\n is positive for any $i$ and any sequence $\bar w$, whose elements
$w_j$ take only the two values: $c$ or $1-c$, but there exists a
sequence $\bar w'$ with only one $0$ element, such that
$\Omega(\bar w',i)<0$ for some $i$. The simplest example is a
one-dimensional chain with symmetrical coupling and $c<1/3$.

\begin{theorem} Let the graph $\Gamma$ does not have ``free'' nodes.
Then all SBR measures are localized around the singular points. If
there exists a ``free'' node, then the localization phenomenon does not
take place. \end{theorem}

Recall, that the definition of the ``free'' node was done in Introduction.

According to this statement one can encode all possible SBR measures by
binary sequences, taking 0 if the corresponding $x_i \le 1/2$
(actually, close to $c$) and $1$ otherwise (close to $1-c$). It turns
out that all the binary sequences can appear here, except ones such that
for some node all its neighbors (in the graph $\Gamma$) have the same value.
For example, for the linear cyclic chain the only restriction for the
encoding sequence is that among 3 consecutive values there should be both 0
and 1.

If instead of the condition above we assume that

\be  \Omega(\bar w,i)>0 \ee

\n for any sequence $\bar w$ with elements $0,c,1-c,1$, then the
localization takes place independently on the existence of ``free''
nodes. Example: a one-dimensional chain with simmetric coupling and
$1/3 < c <1/2$.

Under the last condition CML is ``superstable'' with respect to
various perturbations in the following sense. Consider a localized
system and let us change the position in only one component, say
$x_i$. It turns out that the binary code, corresponding to the
perturbed map will be changed only in this place, and under the
action of the dynamics the code will converge to the original one.

Influence of random noise. Let us take instead of one of the nodes
of our network a source of independent random noise. Then the whole
system will be stable (unstable) with respect to this noise exactly
as in the case of chaotic maps.

%--------------------------------------------------------

\end{document}